\documentclass{article}

\usepackage{obs_study_style}
\usepackage{enumerate}
\usepackage{comment}
\usepackage{todonotes}
\usepackage{subcaption}

\newcommand{\E}{\mathbb{E}}
\newcommand{\Dt}{\tilde D}

\newcommand*\samethanks[1][\value{footnote}]{\footnotemark[#1]}

\begin{document}
	
	\title{Causaltoolbox---Estimator Stability for Heterogeneous Treatment Effects}
	
	\author{\name S\"oren R. K\"unzel\thanks{These authors contributed equally to this work.} 
	\email srk@berkeley.edu\\
		\addr Department of Statistics\\
		University of California, Berkeley 
		\AND
		\name Simon J. S. Walter\samethanks 
		\email sjswalter@berkeley.edu\\
		\addr Department of Statistics\\
		University of California, Berkeley
		\AND
		\name Jasjeet S. Sekhon 
		\email sekhon@berkeley.edu\\
		\addr Department of Political Science \& Department of Statistics\\
		University of California, Berkeley
		}
	
	\maketitle
	
	\begin{abstract}
		Estimating heterogeneous treatment effects has become increasingly important in many fields and life and death decisions are now based on these estimates: for example selecting a personalized course of medical treatment. 
Recently, a variety of procedures relying on different assumptions have been suggested  for estimating heterogeneous treatment effects.
Unfortunately, there are no compelling approaches that allow identification of the procedure that has assumptions that hew closest to the process generating the data set under study and researchers often select one arbitrarily. This approach risks making inferences that rely on incorrect assumptions and gives the experimenter too much scope for $p$-hacking.  A single estimator will also tend to overlook patterns other estimators could have picked up.
We believe that the conclusion of many published papers might change had a different estimator been chosen and we suggest that practitioners should evaluate many estimators and assess their similarity when investigating heterogeneous treatment effects.  
We demonstrate this by applying 28 different estimation procedures to an emulated observational data set; this analysis shows that different estimation procedures may give starkly different estimates. 
We also provide an extensible \texttt{R} package which makes it straightforward for practitioners to follow our recommendations.
	\end{abstract}
	

\section{Introduction} \label{sec:intro}
Heterogeneous Treatment Effect (HTE) estimation is now a mainstay in many disciplines, including personalized medicine \citep{Henderson2016,powers2018some}, 
digital experimentation \citep{taddy2016nonparametric},
economics \citep{Athey2016}, 
political science \citep{green2012modeling},  and
statistics \citep{tian2014simple}. Its prominence has been driven by the combination of the rise of big data, which permits the estimation of fine-grained heterogeneity, and recognition that many interventions have heterogeneous effects, suggesting that much can be gained by targeting only the individuals likely to experience the most positive response. This increase in interest amongst applied statisticians has been accompanied by a burgeoning methodological and theoretical literature: there are now many methods to characterize and estimate  heterogeneity; some recent examples include \cite{hill2011bayesian, athey2015machine, kunzel2017meta, wager2015estimation, nie2017learning}. Many of these methods are accompanied by  guarantees suggesting they possess desirable properties when specific assumptions are met, however, verifying these assumptions may be impossible in many applications; so practitioners are given little guidance for choosing the best estimator for a particular data set. As an alternative to verifying these assumptions we suggest practitioners construct a large family of HTE estimators and consider their similarities and differences. 

Treatment effect estimation contrasts with prediction problems, where researchers can  use cross-validation (CV) or a validation set to compare the performance of different estimators or to combine them in an ensemble. This is infeasible for  treatment effect estimation because of the fundamental problem of causal inference: we can never observe the treatment effect for any individual unit directly, so we have no source of truth to validate or cross-validate against.
Partial progress has been made in addressing this problem; for example, \cite{athey2015machine} suggest using the transformed outcome as the truth, a quantity equal in expectation to the individual treatment effect and \cite{kunzel2017meta} suggests using matching to impute a quantity similar to the unobserved potential outcome. 
However, even if there were a reliable procedure for identifying the estimator with the best predictive performance, we maintain that using multiple estimates can still be superior, because the best performing method or ensemble of methods may perform well in some regions of the feature space and badly in others; using many estimates simultaneously may permit identification of this phenomenon.
For example, researchers can construct a worst-case estimator that is equal to the most pessimistic point estimate, for each point in the feature space, or they can use the idea of stability \citep{yu2013stability}
to assess whether one can trust estimates for a particular subset of units.


\section{Methods} 

\subsection{Study setting}
The data set we analyzed was constructed for the Empirical Investigation of Methods for Heterogeneity Workshop at the 2018 Atlantic Causal Inference Conference. The organizers of the workshop: Carlos Carvalho, Jennifer Hill, Jared Murray, and Avi Feller used the National Study of Learning Mindsets, a randomized controlled trial in a probability sample of U.S. public high schools, to simulate an observational study. The organizers did not disclose how the simulated observational data were derived from the experimental data because the workshop was intended to evaluate procedures for analyzing observational studies, where the mechanism of treatment assignment is not known \em a priori\em . 

\subsection{Measured variables}
The outcome was a measure of student achievement; the treatment was the completion of online exercises designed to foster a learning mindset. Eleven covariates were available for each student:
four are specific to the student and describe the self-reported expectations for success in the future, race, gender and whether the student is the first in the family to go to college;
the remaining seven variables describe the school the student is attending measuring urbanicity, poverty concentration, racial composition, the number of pupils, average student performance, and the extent to which students at the school had fixed mindsets; an anonymized school id recorded which students went to the same school. 

\subsection{Notation and estimands}
For each student, indexed by $i$, we observed a continuous outcome, $Y_i$, a treatment indicator variable, $Z_i$, that is $1$ if the student was in the treatment group and $0$ if she was in the control group, and a feature vector $X_i$. We adopt the notation of the Neyman-Rubin causal model: for each student we assume there exist two potential outcomes: if a student is assigned to treatment we observe the outcome  $Y_i = Y_i(1)$ and if the student is assigned to control we observe $Y_i = Y_i(0)$. Our task was to assess whether the treatment was effective and, if so whether the effect is heterogeneous. In particular, we are interested in discerning if there is a subset of units for which the treatment effect is particularly large or small. 

To assess whether the treatment is effective, we considered the average treatment effect, 
$$
\mbox{ATE} := \E[Y_i(1) - Y_i(0)],
$$
and to analyze the heterogeneity of the data, we computed average treatment effects for a selected subgroup $S$, 
$$
\E[Y_i(1) - Y_i(0) | X_i \in S],
$$
and the Conditional Average Treatment Effect (CATE) function,
$$
\tau(x) := \E [Y_i(1) - Y_i(0)| X_i = x]. 
$$

\subsection{Estimating average effects}
Wherever we computed the ATE or the ATE for some subset, we used four  estimators. 
Three of which were based on the \texttt{CausalGAM} package of \cite{causalGAM}. This package uses generalized additive models to estimate the expected potential outcomes, 
$\hat{\mu}_0(x) := \hat{\E}[Y_i(0) |X_i = x]$, 
and 
$\hat{\mu}_1(x) := \hat{\E}[Y_i(1) |X_i = x]$, 
and the propensity score: 
$\hat{e}(x) := \hat{\E}[Z_i|X_i = x]$. 
With these estimates we computed the Inverse Probability Weighting (IPW) estimator,
$$\hat{\mbox{ATE}}_{\small{\textrm{IPW}}} := \frac{1}{n}\sum_{i=1}^n \left(\frac{Y_i Z_i}{\hat{e}_i}- \frac{Y_i( 1 - Z_i)}{1-\hat{e}_i} \right),$$
the regression estimator,
$$\hat{\mbox{ATE}}_{\mbox{\small{Reg}}} := \frac{1}{n}\sum_{i=1}^n \left[\hat{\mu}_1(X_i) - \hat{\mu}_0(X_i) \right], $$
and the Augmented Inverse Probability Weighted (AIPW) estimator,
$$\hat{\mbox{ATE}}_{\small{\textrm{AIPW}}} := \frac{1}{2n}\sum_{i=1}^n \left(\frac{[Y_i  - \hat{\mu}_0(X_i)]Z_i}{\hat{e}_i}+ \frac{[\hat{\mu}_1(X_i) - Y_i][ 1 - Z_i]}{1-\hat{e}_i} \right).$$ 

We also used the \texttt{Matching} package of \cite{matching2011sekhon} to construct a matching estimator for the ATE.
Matches were required to attend the same school as the student to which they were matched and be assigned to the opposite treatment status. Among possible matches satisfying these criteria we selected the student minimizing the Mahalonobis distance on the four student specific features.

\subsection{Characterizing heterogeneous treatment effects}
\begin{figure}
	\centering
	\includegraphics[width=.9\linewidth]{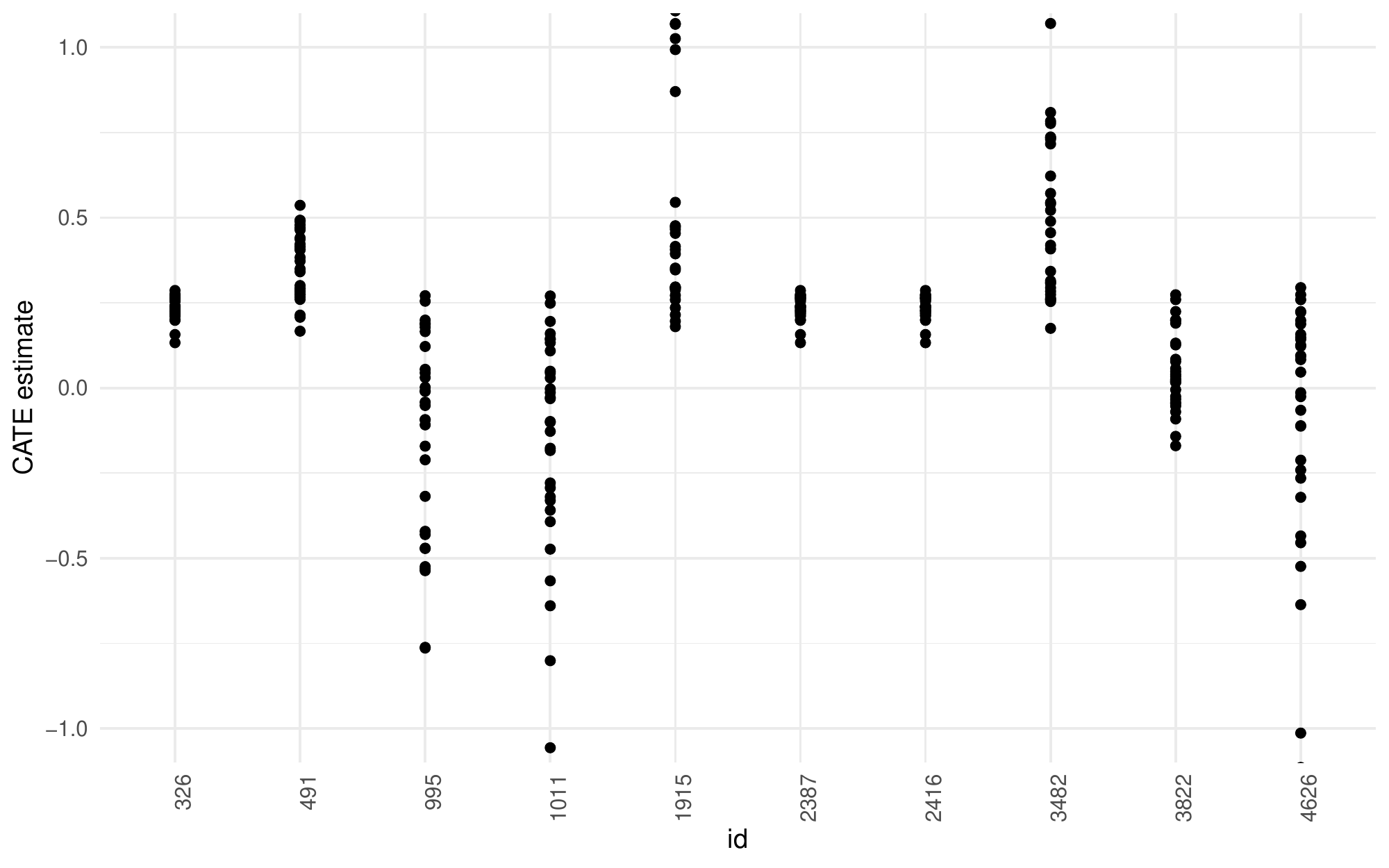}
	\caption{CATE estimation for ten units. For each unit, the CATE is estimated using 28 different estimators.}
	\label{fig:individualDisagreement}
\end{figure}

In any data set there might be some units where estimators significantly disagree; when this happens, we should not trust any estimate unless we understand why certain estimates are unreasonable for these units\footnote{A standard way to capture estimation uncertainty is to report the standard errors or confidence intervals of a single estimator, and we recommend using this approach as well. However, such  methods can be misleading and should not be trusted blindly. For example, in Appendix C of \cite{kunzel2017meta}, the authors found that in regions without overlap, bootstrap confidence intervals were smaller than in regions with overlap. The confidence intervals suggested that in regions without overlap the estimates were more trustworthy, while the opposite was was true.}. Instead of simply reporting an estimate that is likely wrong, we should acknowledge that a conclusions cannot be drawn and more data or domain specific knowledge is needed. Figure \ref{fig:individualDisagreement} demonstrates this phenomenon arising in practice. It shows the estimated treatment effect for ten subjects corresponding to 28 CATE estimators (these estimates arise from the data analyzed in the remainder of this paper). Some of these estimators may have better generalization error than others. However, a reasonable analyst could have selected any one of them. 
We can see that for five units the estimators all fall in a tight cluster, but for the remaining units, the estimators disagree markedly. This may be due to those units being in regions with little overlap, where the estimators overcome data scarcity by pooling information in different ways.

In this analysis, our goal was to understand and interpret the heterogeneity that exists in the treatment response.
An estimate of the CATE function describes the heterogeneity. However, this estimate is hard to interpret, and drawing statistically significant conclusions based on it is difficult. Therefore, we sought large subgroups with a markedly different average treatment effects to help characterize the heterogeneity.

Specifically, we split the data into an \textbf{exploration set} and an equally sized \textbf{validation set}. 
We used the exploration set to identify subsets, which may have a very different behavior from the rest of the students. To do this, we used all CATE estimators trained on the exploration set and we carefully formulated hypotheses based on plots of all the CATE estimates: For example, based on plots of the CATE estimates we might theorize that students in schools with more than 900 students have a much higher treatment effect than those in schools with less than 300 students. 
Next, we used the validation set to verify our findings by estimating the ATEs of each of the subgroups. 

The exploration and validation sets were constructed so that all students from the same school are in the same set: we do not randomize on student level, but on school-level. 
This is important, because it mirrors the probability sampling approach used to construct the full sample; it also means that we can argue that the estimand captured by evaluating our hypotheses on the validation set is the estimand corresponding to the population from which all schools were drawn. 


\subsection{CATE estimators} \label{section:CateEstimators}
We use several procedures to estimate the CATE and we give a brief overview of the procedures here; however, interested readers should consult the referenced papers for a complete exposition.

Many of the procedures can be classified as meta-learners: they modify base learners designed for standard non-causal prediction problems to estimate the CATE.  This is advantageous because we can select a base learner that performs well on the observed data. 

\begin{enumerate}
	\item 
	The \textbf{T-Learner} is the most common meta-learner. Base learners are used to estimate the control and treatment response function separately, $\hat\mu_1(x) := \hat\E[Y_i(1) |X_i =x]$ and $\hat\mu_0(x) := \hat\E[Y_i(0)|X = x]$. The CATE estimate is then the difference between these two estimates, $\hat \tau^T(x) := \hat \mu_1(x) - \hat \mu_0(x)$.
	\item 
	The \textbf{S-Learner} uses one base learner  to estimate the joint outcome function, $\hat\mu(x, z) := \hat\E[Y_i| X_i= x, Z_i = z]$. The predicted CATE is the difference between the predicted values when the treatment assignment indicator is changed from treatment to control, $\hat \tau^S(x) := \hat \mu(x, 1) - \hat \mu(x, 0)$.
	\item 
	The \textbf{MO-Learner} \citep{rubin2007doubly, Walter2017} is a two stage meta-learner. It first uses the base learners to estimate the propensity score, $\hat e(x) := \hat\E[Z_i | X = x]$, and the control and treatment response functions. 
	It then defines the  adjusted modified outcome as 
	$$
	R_i := \frac{Z_i - \hat e(x_i)}{\hat e(x_i)[1 - \hat e(x_i)]} 
	\Big(Y_i - \hat \mu_1(x_i) [1 - \hat e(x_i)] - \hat \mu_0(x_i)\hat e(x_i)\Big).
	$$
	An estimate of the CATE is obtained by using a base learner to estimate the conditional expectation of $R_i$ given $X_i$, $\hat \tau^{MO}(x) := \hat \E[R_i | X_i = x]$. 
	\item 
	The \textbf{X-Learner} \citep{kunzel2017meta} also uses base learners to estimate the response functions and the propensity score. 
	It then defines the imputed treatment effects for the treatment group and control group separately as $\Dt^1_{i} := Y_{i}(1) - \hat \mu_0(X_{i})$ and $\Dt^0_{i}:= \hat \mu_1(X_{i}) - Y_{i}(0)$.
	The two estimators for the CATE are obtained by using base learners to estimate the conditional expectation of the imputed treatment effects, 
	$\hat \tau^X_1 := \hat \E[\Dt^1_i | X_i = x]$, and $\hat \tau^X_0 := \hat \E[\Dt^0_i | X_i = x]$.
	The final estimate is then  a convex combination of these two estimators, 
	$$\hat \tau^X(x)  := \hat e (x) \hat \tau^X_0(x) + (1 - \hat e(x))    \hat \tau^X_1(x).$$
\end{enumerate}

All of these meta-learners have different strength and weaknesses. For example, the T-Learner performs particularly well when the control and treatment response function are simpler than the CATE. The S-Learner performs particularly well when the expected treatment effect is mostly zero or constant.
The X-Learner, on the other hand, has very desirable properties when either the treatment or control group is much larger than the other group.

Note, however, that  all of these meta-learners need a base learners to be fully defined. We believe that tree-based estimators perform well on mostly discrete and low-dimensional data sets. Therefore, we use the \texttt{causalToolbox} package \citep{causalToolbox} that implements all of these estimators combined with RF and BART. 

Using two different tree estimators is desirable because CATE estimators based on BART perform very well when the data-generating process has some global structure (e.g., global sparsity or linearity), while random forest is better when the data has  local structure that does not necessarily generalize to the entire space.   
However, to protect our analysis from biases caused by using tree-based approaches only, we also included methods based on neural networks. We followed \cite{kunzel2018transfer} and implemented the S, T and X-Neural Network methods.

We also included non-meta-learners that are  tree-based, and we believe would work well on this data set:
\begin{enumerate}
	\setcounter{enumi}{4}
	\item The \textbf{causal forest} algorithm \citep{wager2017estimation} is a generalization of the random forest algorithm to estimates the CATE directly. Similar to random forest, it is an ensemble of many tree estimators. Each of the tree estimators follows a greedy splitting strategy to generate leaves for which the CATE function is as homogeneous as possible. The final estimate for each tree for a unit with features $x$ is the difference-in-means estimate of all units in the training set that fall in the same leaf as $x$.  
	\item The \textbf{R-Learner} \citep{nie2017learning} is a set of algorithms that use an approximation of the following optimization problem to estimate the CATE, 
	$$
	\arg\min_\tau \left\{  \frac1n \sum^n_{i = 1} 
	\bigg( 
	\left(Y_i -\hat \mu^{(-i)}(X_i)\right) 
	- 
	\left(W_i - \hat e^{(-i)}(X_i)\right) \tau(X_i) 
	\bigg)^2 + 
	\Lambda_n(\tau(\cdot))\right\}.
	$$
	$\Lambda_n(\tau(\cdot))$ is a regularizer 
	and 
	$\mu^{(-i)}(x)$ and $\hat e^{(-i)}(X_i)$ are held-out predictions of  $\mu(x) = \E[Y_i | X_i = x]$ and the propensity score, $e(x)$, respectively. 
	There are several versions of the R-Learner; we have decided to use one that is based on XGBoost \citep{Chen:2016:XST:2939672.2939785} and one that is based on RF.
\end{enumerate}



 Although we expected there would be school-level effects, and that both the expected performance of each student and the CATE would vary from school to school, it was not clear how to incorporate the school id. 
The two choices we considered were to include a categorical variable recording the school id, or to ignore it entirely. The former makes parameters associated with the six school-level features essentially uninterpretable because they cannot be identified separately from the school id; the second may lead to less efficient estimates because we are denying our estimation procedure the use of all data that was available to us.  
Because we do not want our inference to depend on this decision, we fit each of our estimators twice, once including school id as a feature and once excluding it. We considered 14 different CATE estimation procedures; since each procedure was applied twice, a total of 28 estimators were computed. 

\section{Workshop Results} \label{sec:workshopResults}
Our sample consisted of about 10,000 students enrolled at 76 different schools. The intervention was applied to  33\% of the students. Pre-treatment features were similar in the treatment and control groups but some statistically significant differences were present. 
Most importantly a variable capturing self-reported expectations for success in the future had mean 5.22 (95\% CI, 5.20-5.25) in the control group and mean 5.36 (5.33-5.40) in the treatment group. This meant students with higher expectations of achievement were more likely to be treated.

We assessed whether overlap held by fitting a propensity score model and we found that propensity score estimate for all students in the study was between 0.15 and 0.46 therefore, the overlap condition is likely to be satisfied.

\subsection{Average treatment effects} \label{sec:ATE}
The IPW, regression, and AIPW estimator yielded estimates identical up to two significant figures: $0.25$ with 95\% bootstrap confidence interval of  $(0.22, 0.27)$. The matching estimator gave a similar ATE estimate of $0.26$ with confidence interval $(0.23, 0.28)$.  

The similarity of all the estimates we evaluated is reassuring, but we cannot exclude the possibility that the experiment is affected by an unobserved confounder that affects all estimators in a similar away. To address this we characterize the extent of hidden bias required to explain away our conclusion. We conducted a sensitivity analysis for the matching estimator using the \texttt{sensetivitymv} package of \cite{rosenbaum2018sensitivity}. 
We found that a permutation test for the matching estimator still finds a significant positive treatment effect provided the ratio of the odds of treatment assignment for the treated unit relative to the odds of treatment assignment for the control unit in each pair can be bounded by $0.40$ and $2.52$.
This bound is not very large, and it is  plausible that there exists an unobserved confounder that increases the treatment assignment probability for some unit by a factor of more than 2.52. More information about the treatment assignment mechanism would be required to conclude whether this extent of confounding exists.

\subsection{Heterogeneous effects}
\begin{figure}[h]
	\begin{subfigure}{.5\textwidth}
		\centering
		\includegraphics[width=1\linewidth]{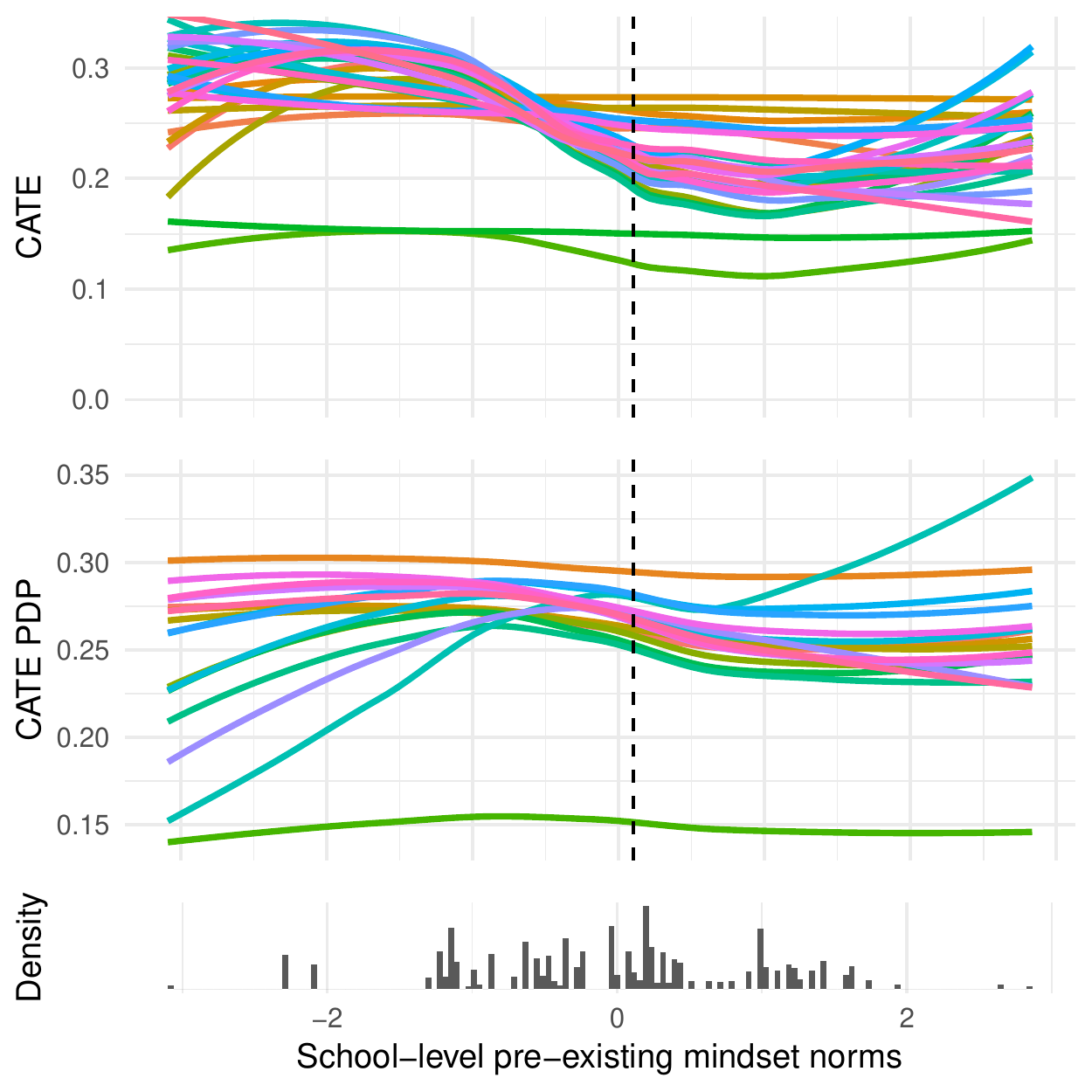}
		\label{fig:sfig1}
	\end{subfigure}%
	\begin{subfigure}{.5\textwidth}
		\centering
		\includegraphics[width=1\linewidth]{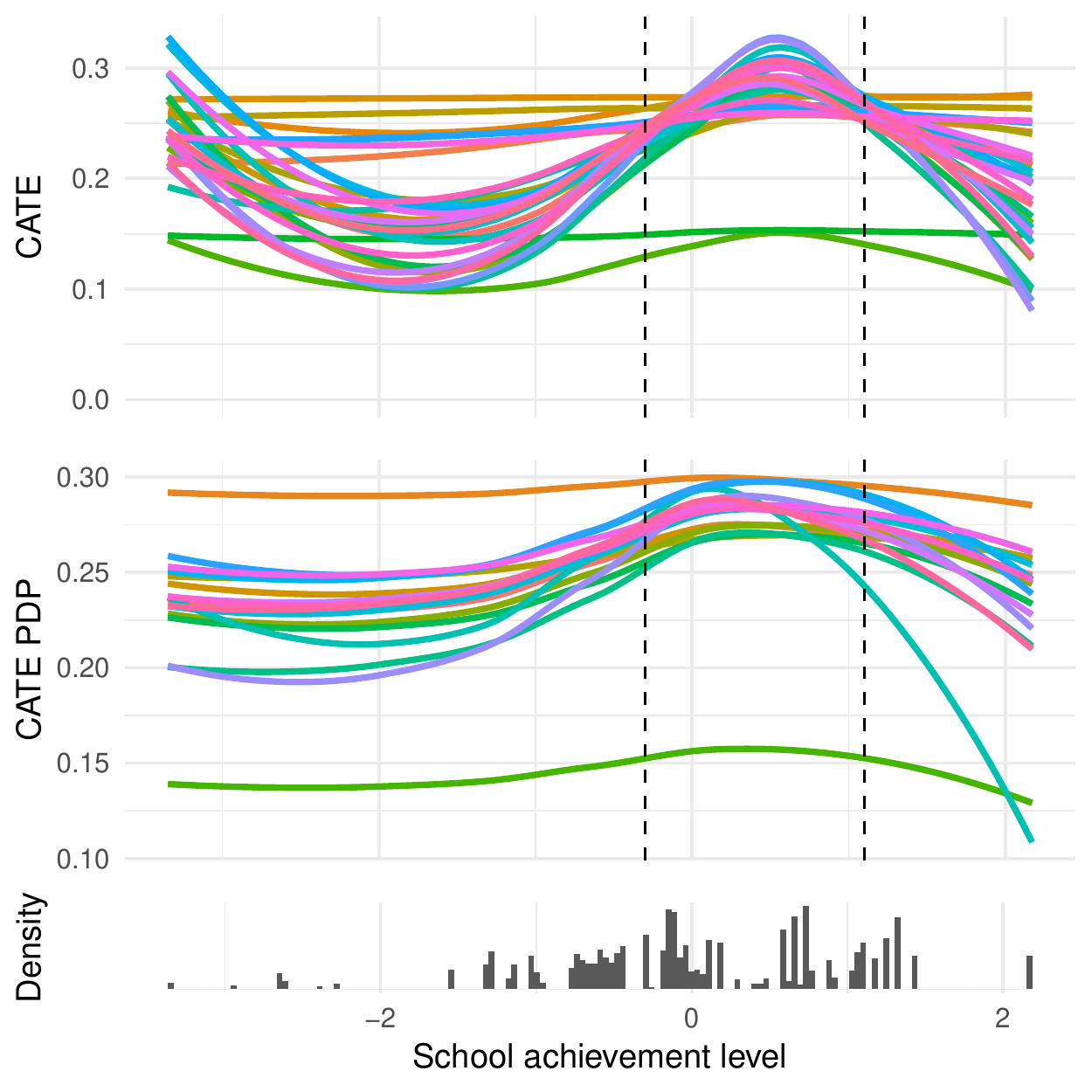}
		\label{fig:sfig2}
	\end{subfigure}
	\caption{Marginal CATE and Partial Dependence Plot (PDP) of the CATE as a function of school-level pre-existing mindset norms and school achiemvent level.}
	\label{fig:continuousHeterogeneity}
\end{figure}

The marginal distribution and partial dependence plots for the 28 CATE estimators as a function of school-level pre-existing mindset norms are shown on the left hand side of Figure \ref{fig:continuousHeterogeneity}. 
There appears to be substantial heterogeneity present: students at schools with mindset norms lower than 0.15 may have a larger treatment effect than students at schools with higher mindset norms. However the Figure suggests the conclusion is not consistent for all of the 28 estimators. 
A similar analysis of the feature recording the school achievement level is shown on the right hand side of this Figure. Again we appear to find the existence of heterogeneity: students with school achievement level near the middle of the range had the most positive response to treatment. On the basis of this figure, we identified thresholds of -0.8 and 1.1 for defining a low achievement level, a middle achievement level, and a high achievement level subgroup. 

We then used the validation set to construct ATE estimates for each of the subgroups.
We found that students who attended schools where the measure of fixed mindsets was less than 0.15 had a higher treatment effect (0.31, 95\% CI 0.26-0.35) than students where the fixed mindset was more pronounced (0.21, 0.17--0.26). Testing for equality of the ATE for these two groups yielded a $p$-value of 0.003.
However, when we considered the subsets defined by school achievement level the differences were not so pronounced. Students at the lowest achieving schools had the smallest ATE estimate 0.19 (0.10--0.32); while students at middle and high-achieving schools had similar ATE estimates: 0.28 (0.24--0.32) and 0.24 (0.16-0.31) respectively. However, none of the pairwise difference between the three groups were significant.

\section{Postworkshop results}
\begin{figure}[h]
	\begin{subfigure}{.5\textwidth}
		\centering
		\includegraphics[width=1\linewidth]{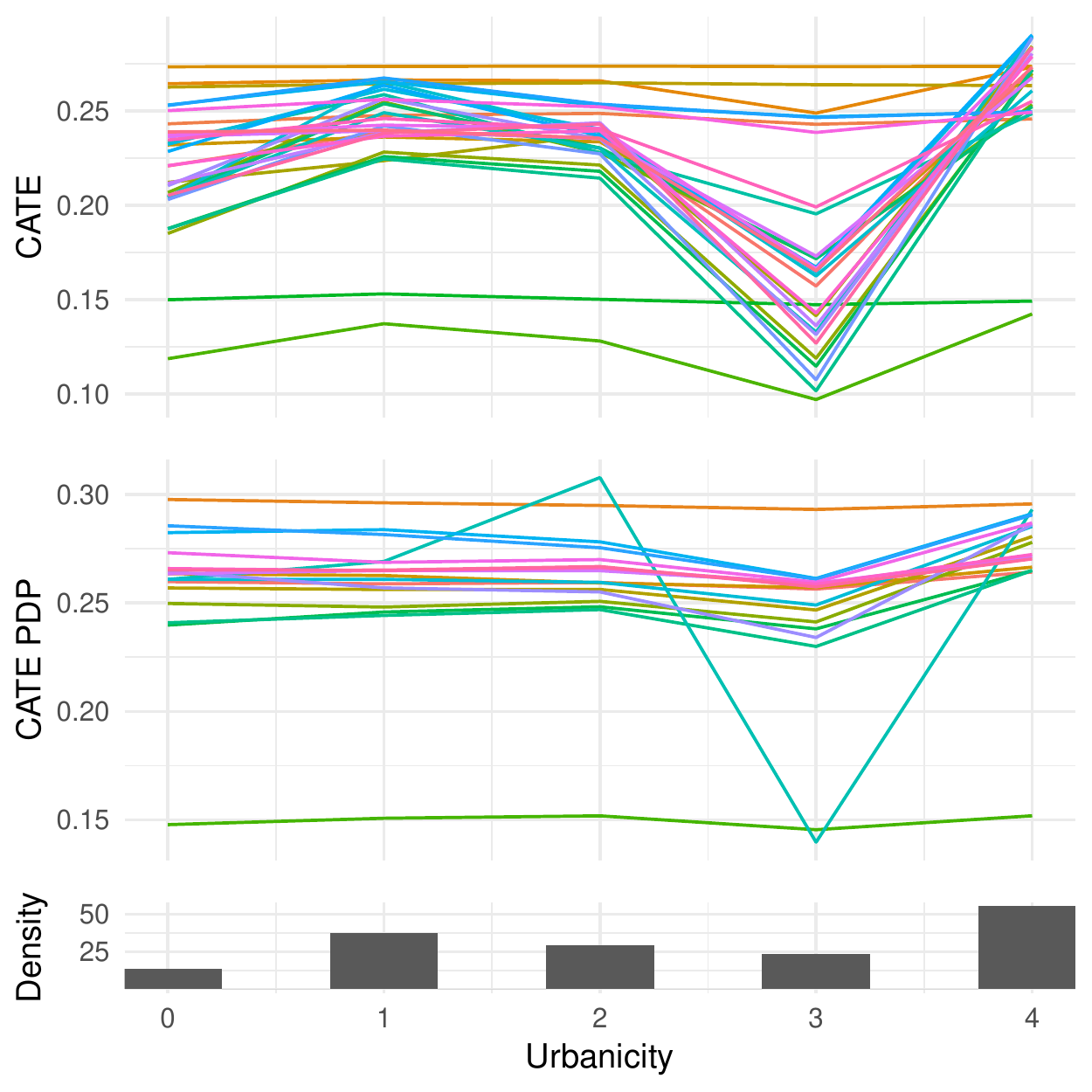}
		\label{fig:discreteH_sfig1}
	\end{subfigure}%
	\begin{subfigure}{.5\textwidth}
		\centering
		\includegraphics[width=1\linewidth]{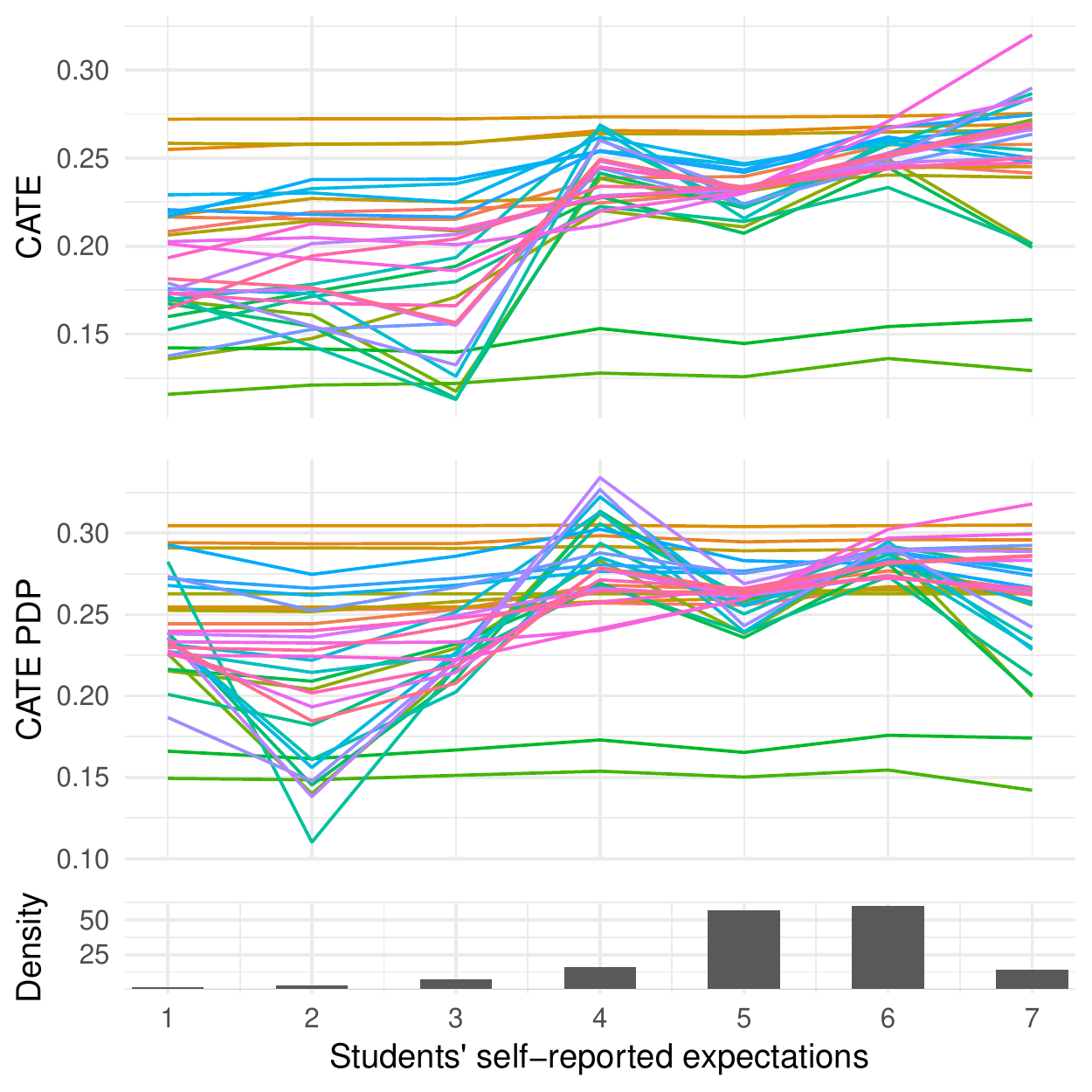}
		\label{fig:discreteH_sfig2}
	\end{subfigure}
	\caption{Marginal CATE and PDP of  Urbanicity and self-reported expectations.}
	\label{fig:discreteHeterogeneityws}
\end{figure}

During the workshop, other contributors found that the variable recording the urbanicity of the schools might explain some of the heterogeneity and we want to analyze this phenomena in the following.
The left hand side of Figure \ref{fig:discreteHeterogeneityws} shows the CATE as a function of urbanicity, and the right hand side of this figure shows the CATE as a function of the student's self-reported expectation of success. 
We formulated two hypothesis: 
students at schools with an urbanicity of 3 seemed to have a lower treatment effect than students at other schools;
students with a self-reported evaluation of 4 might enjoy a higher treatment effect.

These hypotheses were obtained by only using the exploration set; to confirm or refute these hypotheses we used the validation set. 
The validation set confirmed the hypothesis that students at schools with an urbanicity of 3 had a lower treatment effect (0.16, 0.08--0.24) compared to students at schools with a different urbanicity (0.28, 0.25--0.31); however we could not reject the null hypothesis of no difference for the subsets identified by the self-reported evaluation measure. 
The urbanicity test yielded a p-values of 0.008 and the self-reported evaluation test yielded a $p$-value of 0.56.

\section{Discussion}

\subsection{The importance of considering multiple estimators}
The results of our analysis confirm that point estimates of the CATE can differ markedly depending on subtle modelling choices; so we confirm that an analyst's discretion may be the deciding factor in whether and what kind of heterogeneity is found. As the methodological literature on heterogeneous treatment effect estimation continues to expand this problem will become more, not less, serious. To facilitate applying many estimation procedures we have authored an R package \texttt{causalToolbox} that provides a uniform interface for constructing many common heterogeneous treatment estimators. The design of the package makes it straightforward to add new estimators as they are proposed and gain currency.  

Differences that arise in our estimation of the CATE function translate directly into suboptimal real world applications of the treatment considered. To see this we propose a thought experiment: suppose we wanted to determine the treatment for a particular student: a natural treatment rule is to allocate her to treatment if her estimated CATE exceeds a small positive threshold or withhold treatment if it is below the threshold. The analyst might select a CATE estimator on the basis of personal preference or prior experience and it is likely that, for some experimental subjects, the choice of estimator will affect the CATE estimated to such an extent that it changes the treatment decision.  This is particularly problematic in studies where analysts have a vested interest in a particular result and are working without a pre-analysis plan, as they should not have discretion to select a procedure that pushes the results in the direction they desire. 
On the other hand, if analysts consider a wide variety of estimators, as we recommend, and if most estimators agree for an individual, we can be confident that our decision for that individual is not a consequence of arbitrary modelling choices. 
Conversely, if some estimators predict a positive and some a negative response, we should reserve judgment for that unit until more conclusive data is available and admit that we do not know what the best treatment decision is.

\subsection{Would we recommend the online exercises?}
We find that the overall effect of the treatment is significant and positive. We were not able to identify a subgroup of units that had significant and negative treatment effect and we would therefore recommend the treatment for every student. 
We are, however, concerned that an unobserved confounder exists. Our sensitivity analysis showed that our findings would still hold if the confounder is not too strong. We cannot exclude the possibility that there is a strong confounder and would have to know more about the assignment mechanism to address this question. This is particularly problematic, because we have seen that students who had higher expectations for success in the future were more likely to be in the treatment group. 
Uncovering the heterogeneity in the CATE function proved to be substantially more difficult. We found heterogeneity could be identified from school-level pre-exising mindset norms and urbanicity but in general we had limited power to detect heterogeneous effects. For example, experts believe that the heterogeneity might be moderated by pre-existing mindset norms and school-level achievement. For both covariates, we see that most CATE estimators produce estimates that are consistent with this theory.
Domain experts also believe that there could be a "Goldilocks effect" where middle-achieving schools have the largest treatment effect. We are not able to verify this statistically, but we do observe that most CATE estimators describe such an effect.

	
	\acks{We thank Carlos Carvalho, Jennifer Hill, Jared Murray, and Avi Feller for organizing the Empirical Investigation of Methods for Heterogeneity Workshop and their valuable feedback. We also thank the Berkeley Institute for Data Science, the Gordon and Betty Moore Foundation, the Alfred P. Sloan Foundation, and the Office of Naval Researh (ONR) grant N00014-15-1-2367.}
	


\newpage
	
	\vskip 0.2in
	\begin{thebibliography}{21}
\providecommand{\natexlab}[1]{#1}
\providecommand{\url}[1]{\texttt{#1}}
\expandafter\ifx\csname urlstyle\endcsname\relax
  \providecommand{\doi}[1]{doi: #1}\else
  \providecommand{\doi}{doi: \begingroup \urlstyle{rm}\Url}\fi

\bibitem[Athey and Imbens(2015)]{athey2015machine}
Susan Athey and Guido~W Imbens.
\newblock Machine learning methods for estimating heterogeneous causal effects.
\newblock \emph{stat}, 1050\penalty0 (5), 2015.

\bibitem[Athey and Imbens(2016)]{Athey2016}
Susan Athey and Guido~W Imbens.
\newblock {Recursive partitioning for heterogeneous causal effects}.
\newblock \emph{Proceedings of the National Academy of Sciences of the United
  States of America}, 113\penalty0 (27):\penalty0 7353--60, 2016.
\newblock ISSN 1091-6490.
\newblock \doi{10.1073/pnas.1510489113}.

\bibitem[Chen and Guestrin(2016)]{Chen:2016:XST:2939672.2939785}
Tianqi Chen and Carlos Guestrin.
\newblock {XGBoost}: A scalable tree boosting system.
\newblock In \emph{Proceedings of the 22nd ACM SIGKDD International Conference
  on Knowledge Discovery and Data Mining}, KDD '16, pages 785--794, New York,
  NY, USA, 2016. ACM.
\newblock ISBN 978-1-4503-4232-2.
\newblock \doi{10.1145/2939672.2939785}.
\newblock URL \url{http://doi.acm.org/10.1145/2939672.2939785}.

\bibitem[Glynn and Quinn(2017)]{causalGAM}
Adam Glynn and Kevin Quinn.
\newblock \emph{CausalGAM: Estimation of Causal Effects with Generalized
  Additive Models}, 2017.
\newblock R package version 0.1-4.

\bibitem[Green and Kern(2012)]{green2012modeling}
Donald~P Green and Holger~L Kern.
\newblock Modeling heterogeneous treatment effects in survey experiments with
  bayesian additive regression trees.
\newblock \emph{Public opinion quarterly}, 76\penalty0 (3):\penalty0 491--511,
  2012.

\bibitem[Henderson et~al.(2016)Henderson, Louis, Wang, and
  Varadhan]{Henderson2016}
Nicholas~C. Henderson, Thomas~A. Louis, Chenguang Wang, and Ravi Varadhan.
\newblock Bayesian analysis of heterogeneous treatment effects for
  patient-centered outcomes research.
\newblock \emph{Health Services and Outcomes Research Methodology}, 16\penalty0
  (4):\penalty0 213--233, Dec 2016.
\newblock ISSN 1572-9400.
\newblock \doi{10.1007/s10742-016-0159-3}.

\bibitem[Hill(2011)]{hill2011bayesian}
Jennifer~L Hill.
\newblock Bayesian nonparametric modeling for causal inference.
\newblock \emph{Journal of Computational and Graphical Statistics}, 20\penalty0
  (1):\penalty0 217--240, 2011.

\bibitem[K{\"u}nzel et~al.(2017)K{\"u}nzel, Sekhon, Bickel, and
  Yu]{kunzel2017meta}
S{\"o}ren K{\"u}nzel, Jasjeet Sekhon, Peter Bickel, and Bin Yu.
\newblock Meta-learners for estimating heterogeneous treatment effects using
  machine learning.
\newblock \emph{arXiv preprint arXiv:1706.03461}, 2017.

\bibitem[K\"unzel et~al.(2018)K\"unzel, Tang, Xie, Saarinen, Bickel, Yu, and
  Sekhon]{causalToolbox}
S\"oren K\"unzel, Allen Tang, Ling Xie, Theo Saarinen, Peter Bickel, Bin Yu,
  and Jasjeet Sekhon.
\newblock \emph{causalToolbox: Toolbox for Causal Inference with emphasize on
  Heterogeneous Treatment Effect Estimator}, 2018.
\newblock R package version 0.0.1.000.

\bibitem[K{\"u}nzel et~al.(2018)K{\"u}nzel, Stadie, Vemuri, Ramakrishnan,
  Sekhon, and Abbeel]{kunzel2018transfer}
S{\"o}ren~R K{\"u}nzel, Bradly~C Stadie, Nikita Vemuri, Varsha Ramakrishnan,
  Jasjeet~S Sekhon, and Pieter Abbeel.
\newblock Transfer learning for estimating causal effects using neural
  networks.
\newblock \emph{arXiv preprint arXiv:1808.07804}, 2018.

\bibitem[Nie and Wager(2017)]{nie2017learning}
Xinkun Nie and Stefan Wager.
\newblock Learning objectives for treatment effect estimation.
\newblock \emph{arXiv preprint arXiv:1712.04912}, 2017.

\bibitem[Powers et~al.(2018)Powers, Qian, Jung, Schuler, Shah, Hastie, and
  Tibshirani]{powers2018some}
Scott Powers, Junyang Qian, Kenneth Jung, Alejandro Schuler, Nigam~H Shah,
  Trevor Hastie, and Robert Tibshirani.
\newblock Some methods for heterogeneous treatment effect estimation in high
  dimensions.
\newblock \emph{Statistics in medicine}, 2018.

\bibitem[Rosenbaum(2018)]{rosenbaum2018sensitivity}
Paul~R. Rosenbaum.
\newblock \emph{sensitivitymv: Sensitivity Analysis in Observational Studies},
  2018.
\newblock R package version 1.4.3.

\bibitem[Rubin and van~der Laan(2007)]{rubin2007doubly}
Daniel Rubin and Mark~J van~der Laan.
\newblock A doubly robust censoring unbiased transformation.
\newblock \emph{The international journal of biostatistics}, 3\penalty0 (1),
  2007.

\bibitem[Sekhon(2011)]{matching2011sekhon}
Jasjeet~S. Sekhon.
\newblock Multivariate and propensity score matching software with automated
  balance optimization: The {Matching} package for {R}.
\newblock \emph{Journal of Statistical Software}, 42\penalty0 (7):\penalty0
  1--52, 2011.

\bibitem[Taddy et~al.(2016)Taddy, Gardner, Chen, and
  Draper]{taddy2016nonparametric}
Matt Taddy, Matt Gardner, Liyun Chen, and David Draper.
\newblock A nonparametric bayesian analysis of heterogenous treatment effects
  in digital experimentation.
\newblock \emph{Journal of Business \& Economic Statistics}, 34\penalty0
  (4):\penalty0 661--672, 2016.

\bibitem[Tian et~al.(2014)Tian, Alizadeh, Gentles, and
  Tibshirani]{tian2014simple}
Lu~Tian, Ash~A Alizadeh, Andrew~J Gentles, and Robert Tibshirani.
\newblock A simple method for estimating interactions between a treatment and a
  large number of covariates.
\newblock \emph{Journal of the American Statistical Association}, 109\penalty0
  (508):\penalty0 1517--1532, 2014.

\bibitem[Wager and Athey(2017{\natexlab{a}})]{wager2015estimation}
Stefan Wager and Susan Athey.
\newblock Estimation and inference of heterogeneous treatment effects using
  random forests.
\newblock \emph{Journal of the American Statistical Association},
  2017{\natexlab{a}}.

\bibitem[Wager and Athey(2017{\natexlab{b}})]{wager2017estimation}
Stefan Wager and Susan Athey.
\newblock Estimation and inference of heterogeneous treatment effects using
  random forests.
\newblock \emph{Journal of the American Statistical Association}, \penalty0
  (just-accepted), 2017{\natexlab{b}}.

\bibitem[Walter et~al.(2018)Walter, Sekhon, and Yu]{Walter2017}
Simon Walter, Jasjeet Sekhon, and Bin Yu.
\newblock Analyzing the modified outcome for heterogeneous treatment effect
  estimation.
\newblock \emph{Unpublished manuscript}, 2018.

\bibitem[Yu(2013)]{yu2013stability}
Bin Yu.
\newblock {Stabiilty}.
\newblock \emph{Bernoulli}, 19:\penalty0 1484--1500, 2013.

\end{thebibliography}
\end{document}